# Ignition limit and shock-to-detonation transition mode of *n*-heptane/air mixture in high-speed wedge flows

Hongbo Guo[a,b], Yong Xu[b], Hongtao Zheng[a], Huangwei Zhang[b,*]

[a] *College of Power and Energy Engineering, Harbin Engineering University Harbin 150001, People's Republic of China*
[b] *Department of Mechanical Engineering, National University of Singapore 9 Engineering Drive 1, Singapore, 117576, Republic of Singapore*

**Abstract**

In this work, oblique detonation of *n*-heptane/air mixture in high-speed wedge flows is simulated by solving the reactive Euler equations with a two-dimensional (2D) configuration. This is a first attempt to model complicated hydrocarbon fuel ODWs with a detailed chemistry (44 species and 112 reactions). Effects of freestream equivalence ratios and velocities are considered, and the abrupt and smooth transition from oblique shock to detonation are predicted. Ignition limit, ODW characteristics, and predictability of the transition mode are discussed. Firstly, homogeneous constant-volume ignition calculations are performed for both fuel-lean and stoichiometric mixtures. The results show that the ignition delay generally increases with the wedge angle. However, a negative wedge angle dependence is observed, due to the negative temperature coefficient effects. The wedge angle range for successful ignition of *n*-heptane/air mixtures decreases when the wedge length is reduced. From 2D simulations of stationary ODWs, the initiation length generally decreases with the freestream equivalence ratio, but the transition length exhibits weakly non-monotonic dependence. Smooth ODW typically occurs for lean conditions (equivalence ratio < 0.4). The interactions between shock / compression waves and chemical reaction inside the induction zone are also studied with the chemical explosive mode analysis. Moreover, the predictability of the shock-to-detonation transition mode is explored through quantifying the relation between ignition delay and chemical excitation time. It is demonstrated that the ignition delay (excitation time) increases (decreases) with the freestream equivalence ratio for the three studied oncoming flow velocities. Smaller excitation time corresponds to stronger pressure waves from the ignition location behind OSW. When the ratio of excitation time to ignition delay is high (e.g., > 0.5 for n-C7H16, > 0.3 for C2H2 and > 0.2 for H2, based on the existing data compilation in this work), smooth transition is more likely to occur.

*Corresponding author. E-mail address:
huangwei.zhang@nus.edu.sg (H. Zhang)

# 1. Introduction

Oblique detonation engine (ODE) is promising since it not only has the advantages of scramjet, but also enjoys high thermal cycle efficiency of detonation combustion [1]. The oblique detonation wave (ODW) in ODE evolves from an oblique shock wave (OSW) attached to a wedge through abrupt or smooth transition [2]. The shock-to-detonation transition is significantly influenced by the interactions between local gas dynamics and chemical reactions behind the OSW [3,4]. Therefore, it is important to understand the mechanism of OSW-ODW transition to design and develop oblique detonation engines.

Pratt et al. [5] obtain the analytical relations between oblique shock/detonation wave angle and inflow conditions by approximating the ODW as an OSW coupled with an instantaneous post-shock heat release. Guo et al. [6] further find that oblique detonation has a large stationary range in fuel-lean methane-air mixtures through thermodynamic theoretical analysis. Bachman et al. [7] study the ignition and stability of oblique detonation waves and propose an accurate ignition criterion of predicting the formation of an ODW for a given inflow temperature, Mach number, wedge angle, and length of the inviscid wedge surface.

The structure of oblique detonation has also been studied. For instance, Li et al. [2] reveal that oblique detonation structure is usually composed of a non-reactive oblique shock wave, an induction zone, a series of deflagration waves, and an oblique detonation wave. Later studies demonstrate that two OSW-ODW transition modes exist. One is smooth, featured by a curved shock, whilst the other is abrupt with a multi-wave point [8,9]. Recent studies by Teng et al. [10] show additional complexities, i.e., the multi-wave point in the abrupt transition zone is connected with compression waves or normal detonation wave.

The initiation and development of oblique detonation are influenced by flow dynamics and chemical kinetics in the induction zone. Figueira da Silva et al. [11] conduct parametric studies on ODW and explore the influences of inflow Mach number, temperature, pressure and wedge angle on ODW transition mode. Zhang et al. [12] find that the dependence of characteristic length on hydrogen-air mixtures equivalence ratios is the U-shaped curve with critical ratio 0.8, and this nonmonotonicity always exists at low pressure and high temperature. However, they do not study their effects on shock-to-detonation transition. Xiang et al. [13] investigate the features of an oblique detonation induced by a finite wedge in the hydrogen-air mixtures with varying equivalence ratios and also observe that the mixtures near the stoichiometric condition have a shorter induction length. The U-shaped dependence is always observed by them for different inflow Mach numbers.

Most previous studies consider hydrogen as the fuel. To the best of our knowledge, limited work is reported in the open literature on ODW with realistic hydrocarbon fuels. Recently, Zhang et al. [14] study the acetylene/oxygen ODWs with argon dilution and their results show that the ODW transition mode is affected by both dilution ratio and freestream speed. However, our understanding about reaction initiation and transition of hydrocarbon fuel ODW are still very limited, particularly about the induction zone chemical structure as well as the mechanism and predictability of the OSW-ODW transition mode. In this work, numerical computations of oblique detonation wave in *n*-heptane/air mixtures will be conducted, through solving the reactive Euler equations with a skeletal mechanism (44 species and 112 reactions) [15]. The objectives of this study are to analyze: (i) ignition limit for stationary *n*-heptane/air oblique detonation, (ii) flow structure and chemical structure of oblique detonations with different transition modes, and (iii) predictability of shock-to-detonation transition mode.

# 2. Numerical method and physical model

## *2.1 Numerical method*

The equations of mass, momentum, energy, and species mass fraction are solved for high-speed *n*-heptane/air wedge flows. Molecular diffusion effects are not considered and hence the wedge boundary layer is not predicted, which is deemed acceptable due to the limited influence on oblique detonations [7]. The equations are solved by an OpenFOAM code, *RYrhoCentralFoam* [16]. Its accuracy in simulating compressible reacting flows has been well confirmed through a range of validations and verifications, e.g., shock capturing, flame-chemistry interaction, detonation speed and cell sizes [16]. Recently, it has been successfully employed for different detonation problems (e.g., [17]).

Cell-centered finite volume method is used for discretizing the above equations. Second-order backward scheme is applied for time marching, and the time step is about $1\times10^{-10}$ second for the ODW modelling. Second-order Godunov-type central and upwind-central scheme [18] is used to discretize the convection terms. The chemical reaction terms are integrated with Euler implicit method, and the result accuracy is comparable to those of other ordinary differential equation solvers [16]. In particular, for *n*-heptane oblique detonations, we compare the Euler implicit method and Rosenbrock ordinary differential equation solver in the supplementary document, and the differences of the ODW structures predicted by them are negligible. A skeletal mechanism with 44 species and 112 reactions [15] is used for *n*-heptane/air combustion. An additional ODW simulation with a more detailed mechanism (88 species and 387 reactions, see the details in supplementary document) is run and it is shown that our results are not sensitive to the chemical mechanisms.

### 2.2 Physical model

Two-dimensional configuration is considered in this study, and the schematic is provided in Fig. 1. A supersonic premixed $n$-$C_7H_{16}$/air gas approaches a semi-infinite wedge with an angle of $\theta$, which is fixed to be $\theta$ = 30° in our CFD simulations. Under appropriate conditions, part of the oblique shock would evolve into ODW and stabilizes in the supersonic flows.

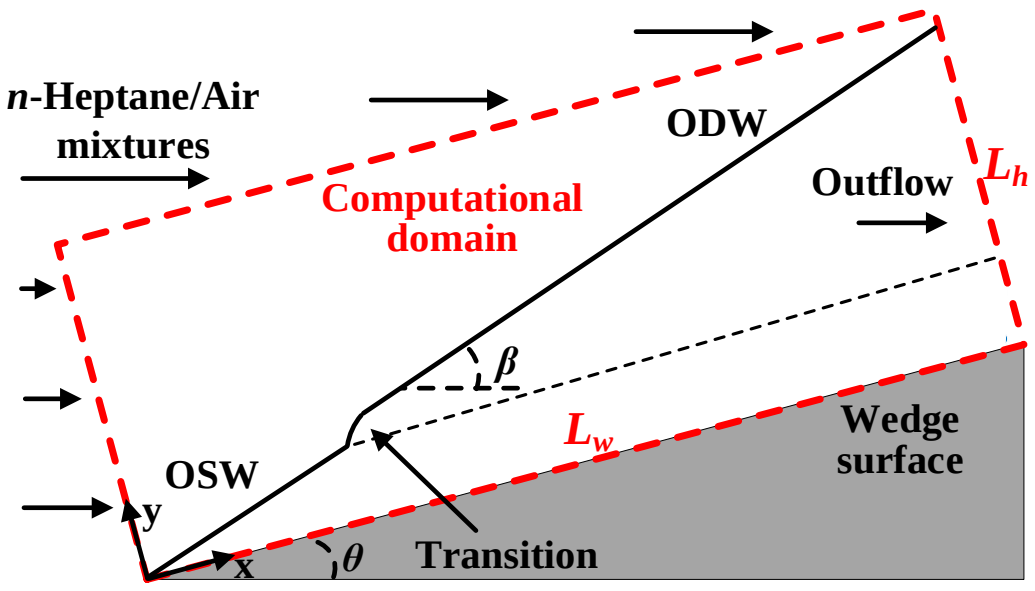

Fig. 1. Schematic of wedge-induced oblique detonation.

Figure 1 shows the computational domain (red box) and boundary conditions. The $x$ ($y$) axis is parallel (normal) to the wedge surface, whilst the origin lies at the leading edge of the wedge. Dirichlet conditions are enforced at the left and top boundaries. The equivalence ratio of the $n$-heptane and air mixtures, $\phi$, is varied from 0.2 to 1.4. Therefore, the species molar ratio follows $n$-$C_7H_{16}/O_2/N_2$ = $\phi$/11/41.36. Moreover, three oncoming flow speeds are considered, i.e., $u_{in}$= 2,800, 2,900 and 3,000 m/s. The corresponding Mach numbers are 8.1–9.1, depending on the gas composition. The freestream temperature and pressure are fixed to be 298 K and 1.0 atm, respectively. Besides, the outlet is non-reflective, whilst the wedge surface is adiabatic and slip wall.

In our simulations, the domain size is adjusted to cover the complete picture of the oblique detonations. The minimum domain size (wedge forebody length $L_w$ × domain height $L_h$) is 0.06 × 0.02 m$^2$, whereas the maximum one is 0.1 × 0.05 m$^2$. Orthogonal grid is used to discretize the domain. The base mesh size is 40 × 40 μm$^2$, based on which one additional level of mesh refinement is made near the OSW and ODW, as well as the areas behind them. This results in a uniform cell size of 20 × 20 μm$^2$ in these locations to sufficiently resolve the detonation flow structures. The total cell number varies from about 1,900,000 to 7,850,000 with various domain sizes. The half-reaction length in Chapman-Jouguet (CJ) detonations with stoichiometric mixtures ($\phi$ = 1.0) is about 1.19 mm. As such, it corresponds to about 59 cells in the half-reaction zone. Mesh sensitivity analysis in the supplementary document shows that further refining the mesh (finest cell with 10 × 10 μm$^2$) does not change the overall ODW structures.

The simulations run on the ASPIRE 1 Cluster from National Supercomputing Centre in Singapore. The simulated physical time varies from 50 to 80 μs in different cases, which is sufficient to capture the stabilized ODWs.

## 3. Results and discussion

### 3.1 Ignition limit

Figure 2 shows the ignition delay $\tau_i$ and initiation length $L_i$ of two $n$-$C_7H_{16}$/air mixtures ($\phi$ = 1.0 and 0.2) as functions of the wedge angles $\theta$. Here $\tau_i$ is determined from zero-dimensional (0D) isochoric autoignition calculations, based on the initial thermodynamic conditions derived from oblique shock relations [6]. It is defined by the instant at which the heat release rate (HRR) reaches the maximum in the homogeneous ignition process. The induction length, $L_i$, is estimated from $L_i = u_s \cdot \tau_i$, where $u_s$ is the gas velocity behind the oblique shock [11]. Note that the ODWs only exist for the solid segment of the curves.

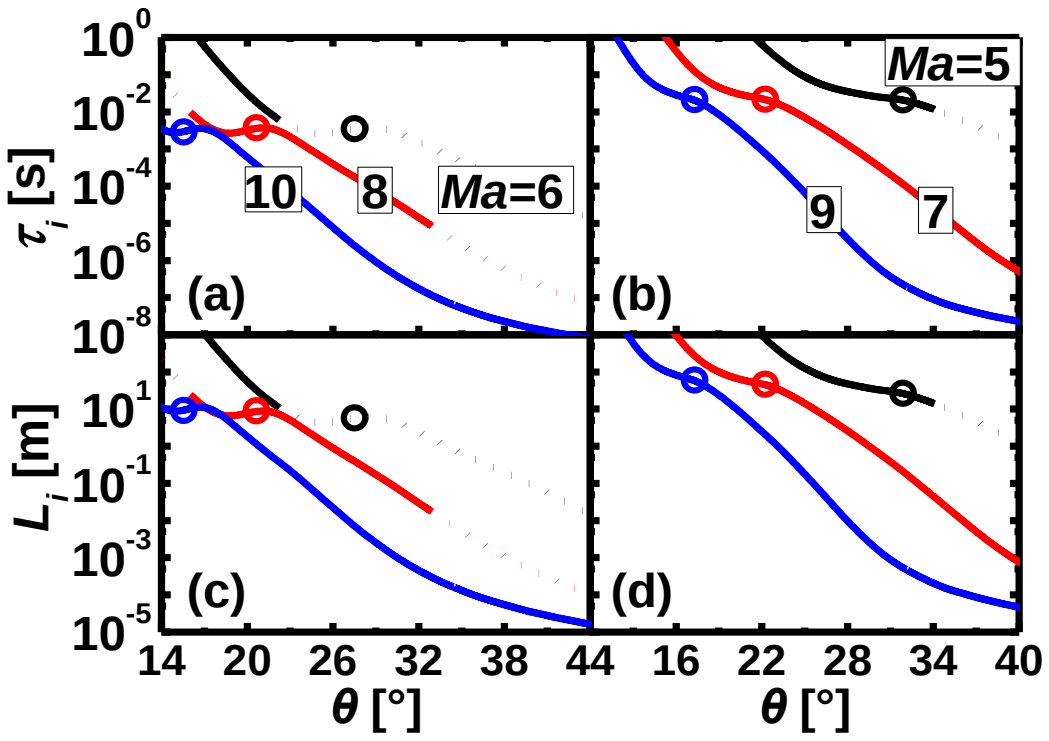

Fig. 2. Ignition delay and induction length as functions of wedge angle with $\phi$ = 1.0 (a and c) and $\phi$ = 0.2 (b and d). Solid line: OSW/ODW; dashed line: OSW only. Each curve left to the circle exhibits multi-stage ignition.

It is seen from Fig. 2 that with increased wedge angle $\theta$, the ignition delay and initiation length generally decrease. This is because the temperature and pressure behind oblique shock increase with $\theta$ (see Fig. S5 in supplementary document), and the induction process is therefore considerably shortened. Multi-stage ignition associated with low-temperature chemistry is observed for the solutions left to the circle for each curve in Fig. 2 (e.g., Fig. S6 in supplementary document). This leads to the opposite $\theta$ dependence for a range of intermediate wedge angles with $\phi$ = 1.0, e.g., 18.8° < $\theta$ < 21.6° when $Ma$ = 8 in Figs. 2(a) and 2(c). Within this range, the temperatures behind the OSW fall in the corresponding range of negative temperature coefficient (NTC) of $n$-$C_7H_{16}$/air mixtures corresponding to the respective post-OSW pressures. The reader should be reminded that along the curves, the pressure and temperature for each data point are different, as seen from Fig. S5. Besides, when Mach number increases from 6 to 10, the ignition delay and initiation length are generally reduced due to higher post-shock temperature and pressure. Moreover, the negative dependence on $\theta$ is not evident in the fuel-lean mixtures, as demonstrated in Figs. 2(b) and 2(d). Overall, their $\tau_i$ and $L_i$ change monotonically with $\theta$.

For the $\phi$ = 1.0 mixtures, $\tau_i$ corresponding to negative $\theta$ dependence is 2.5 − 3.8 ms, whilst the corresponding $L_i$ is 5−10 m. They are smaller than the counterpart results (i.e., the section with flat slope of the $\tau_i - \theta$ curves) with $\phi$ = 0.2, which are $\tau_i$ = 20−40 ms and $L_i$ = 30−70 m, respectively. Based on the existing ODW experiments, for the purpose of the miniaturization of the propulsion system, the used wedge surface is much shorter than the induction length affected by negative $\theta$ dependence. For example, in a recent hydrogen ODW experiment [19], the wedge length is 0.26 m for an abrupt ODW and 0.41 m for a smooth ODW. It is even smaller in the tests by Morris [20]. Despite this, the finding of non-monotonic $\theta$ dependence in Fig. 2 is still valuable. Further experimental/numerical explorations are merited about how this phenomenon affects hydrocarbon ODW dynamics and ODE performance.

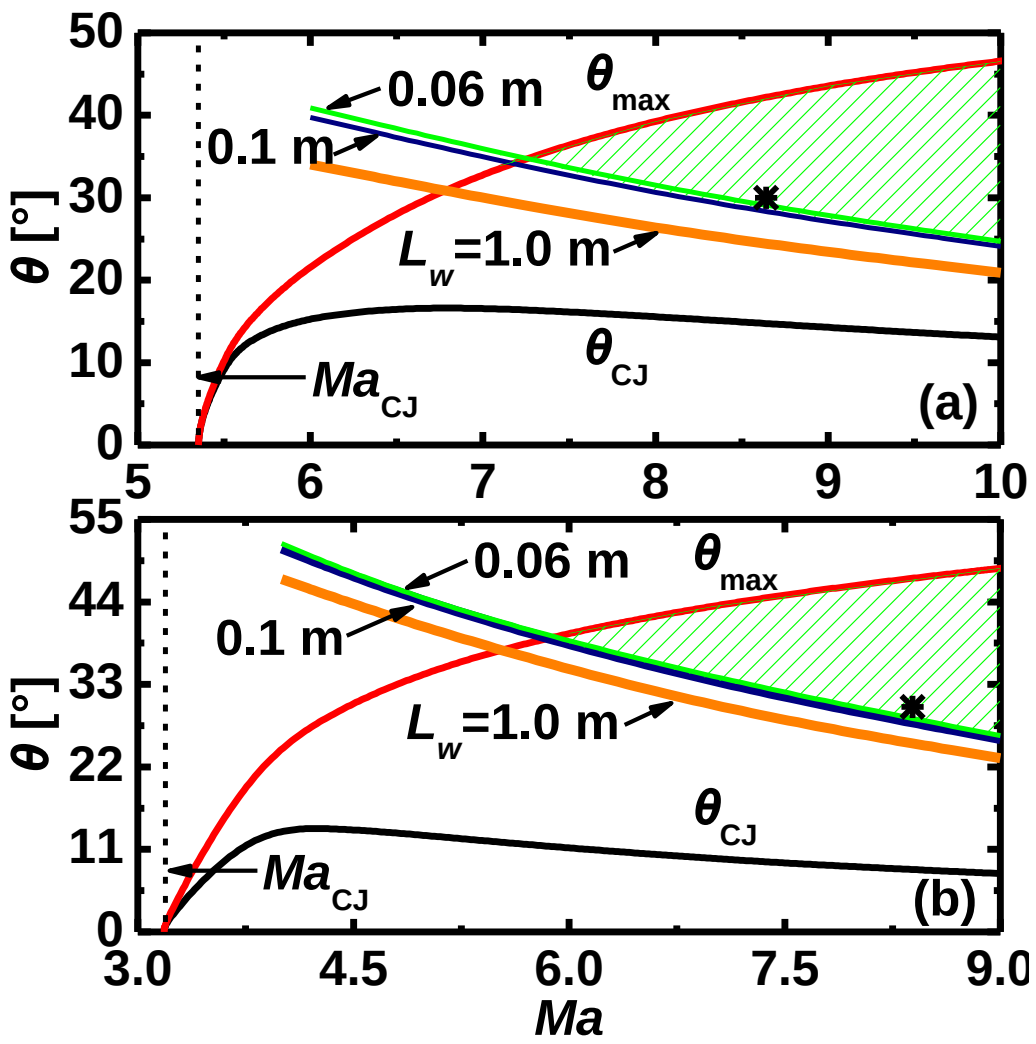

Fig. 3. ODW standing window and ignition limit for different wedge lengths for (a) $\phi$ = 1.0 and (b) $\phi$ = 0.2. Asterisk: CFD cases studied in Section 3.2.

Figure 3 shows the oblique detonation standing windows when the equivalence ratios of $n$-$C_7H_{16}$/air mixtures are 1.0 and 0.2. The CJ Mach numbers $Ma_{CJ}$ (dashed lines) are 5.35 and 3.18 respectively in two mixtures, and only with $Ma > Ma_{CJ}$ the steady solutions for ODW are possible [5]. With increased freestream Mach number $Ma$, the wedge angle range with stable ODWs, i.e., $\theta_{max} < \theta < \theta_{CJ}$, becomes increased. Here $\theta_{max}$ corresponds to the maximum wedge angle, beyond which the ODW is detached, whereas $\theta_{CJ}$ is the angle at which a CJ ODW can stand [5]. Besides, Fig. 3 also shows the lines of critical wedge angle, above which ignition of the combustible gas can proceed over the wedge forebody surface of a specific length, e.g., $L_w$ = 0.06, 0.1 and 1.0 m. They are determined from the results in Fig. 2. When $L_i$ is less than a given wedge length, ignition is deemed successful. In Fig. 3, the area between the critical line for a $L_w$ and $\theta_{max}$ indicates the range of wedge angle with which the corresponding mixture can be ignited over the wedge surface. For instance, the green zone is for $L_w$ = 0.06 m (which is the studied length in the following CFD analysis). As one can see from Fig. 3, the shorter the wedge length (e.g., from 1.0 to 0.06 m), the smaller the wedge angle to achieve successful ignition of combustible gas behind the shock, with higher freestream Mach numbers needed.

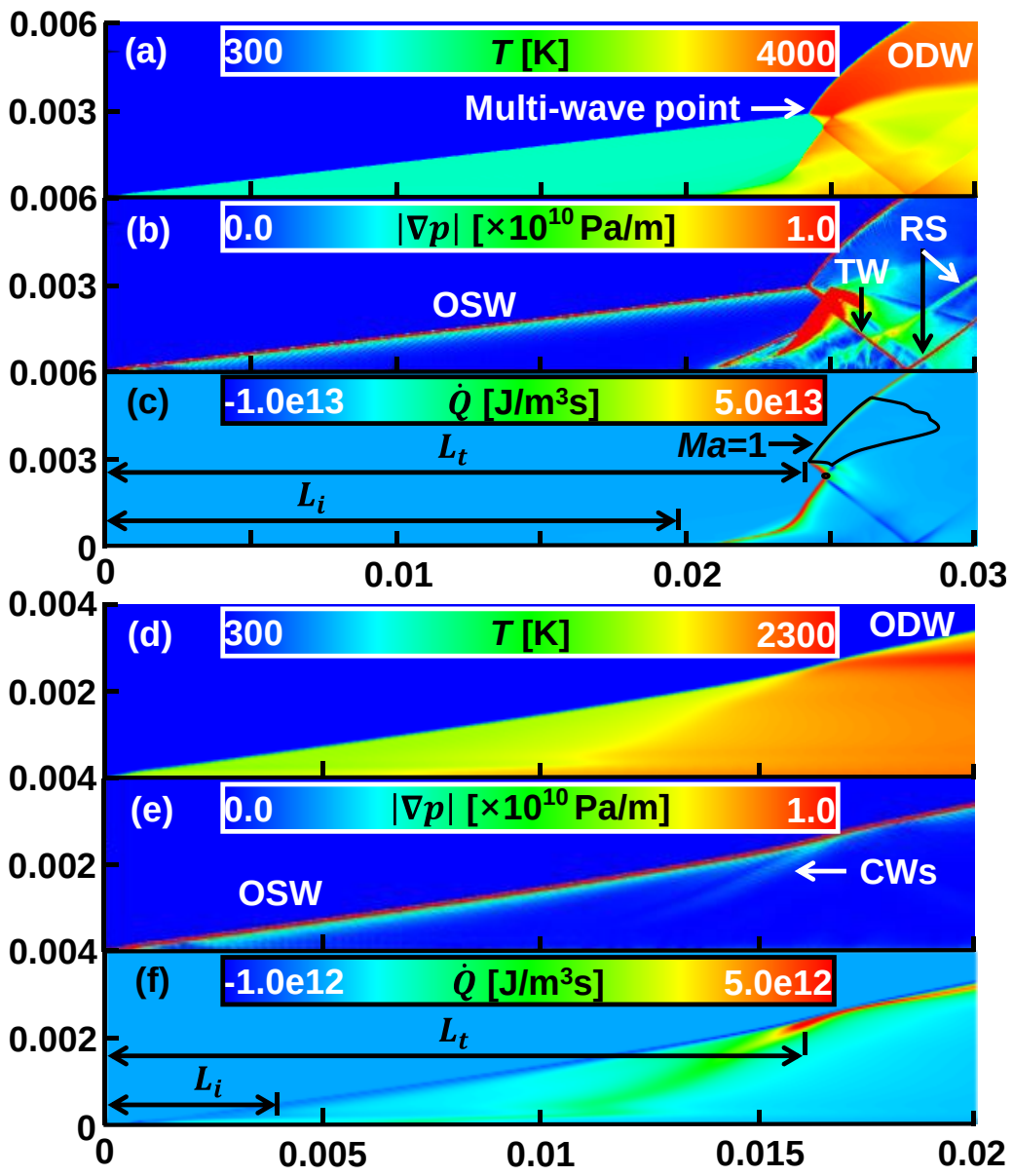

Fig. 4. Temperature, pressure gradient magnitude, and HRR of oblique detonations: (a) – (c) $\phi$ = 1.0; (d) – (f) $\phi$ = 0.2. $u_{in}$ = 2,900 m/s. Axis label unit: m.

### 3.2 *Oblique detonation general characteristics*

In this section, we will focus on two oblique detonations from the ignitability zone for $L_w$ = 0.06 m, marked as asterisks ($\theta$ = 30°) in Figs. 3(a) and 3(b). The freestream conditions: $u_{in}$ = 2,900 m/s, $\phi$ = 1.0 and 0.2. Their corresponding Mach numbers are $Ma$ = 8.64 and 8.39, respectively. Multi-stage ignition of $n$-$C_7H_{16}$/air gas is not present under these conditions.

The distributions of temperature, pressure gradient magnitude and HRR of two oblique detonations are shown in Fig. 4. Note that the transition length $L_t$ is defined as the distance ($x$-direction) from the wedge tip to the OSW-ODW transition locus, while the reaction initiation length $L_i$ is defined as the distance from the wedge tip to the position at which the HRR peaks along the wedge surface (consistent with the definition of induction length in Section 3.1). For the stoichiometric case in Figs. 4(a)−4(c), the mixture is ignited near the wedge surface at an initiation length of $L_i$ = 0.0198 m (marked in Fig. 4c). Strong shock is emanated from the ignition point and travels upward / downstream at a local Mach angle. Mutual enhancement occurs between the shock and local chemical reaction in the induction zone, leading to the formation of a CJ oblique detonation before the transition locus. An abrupt OSW-ODW transition occurs with a salient multi-wave point ($L_t$ = 0.0243 m). A $\lambda$-shaped wave structure can be found, which is also observed in hydrogen ODWs [10]. A local subsonic zone is observed near the multi-wave point due to the local overdriven detonation near the multi-wave point.

For the $\phi$ = 0.2 mixture, the initiation length is $L_i$ = 0.00347 m, much smaller than that of the $\phi$ = 1.0 case. This is because of the higher post-shock temperature (about 1601 K) in this case than that in the stoichiometric case (1452 K, see Fig. S7 of supplementary document). The compression waves (CW in Fig. 4e) from the near-wedge ignition locus are much weaker than the shocks in the $\phi$ = 1.0 case, through comparing the pressure gradient magnitude. How the gas composition affects the CW / SW intensity will be further quantified in Section 3.4. They intersect the precursor shock at approximately 0.0167 m, and the latter becomes curved and smoothly transitions into an ODW through interacting with the reactions behind it (therefore $L_t$ = 0.0167 m). In this case, the flow speeds behind the OSW and ODW are supersonic.

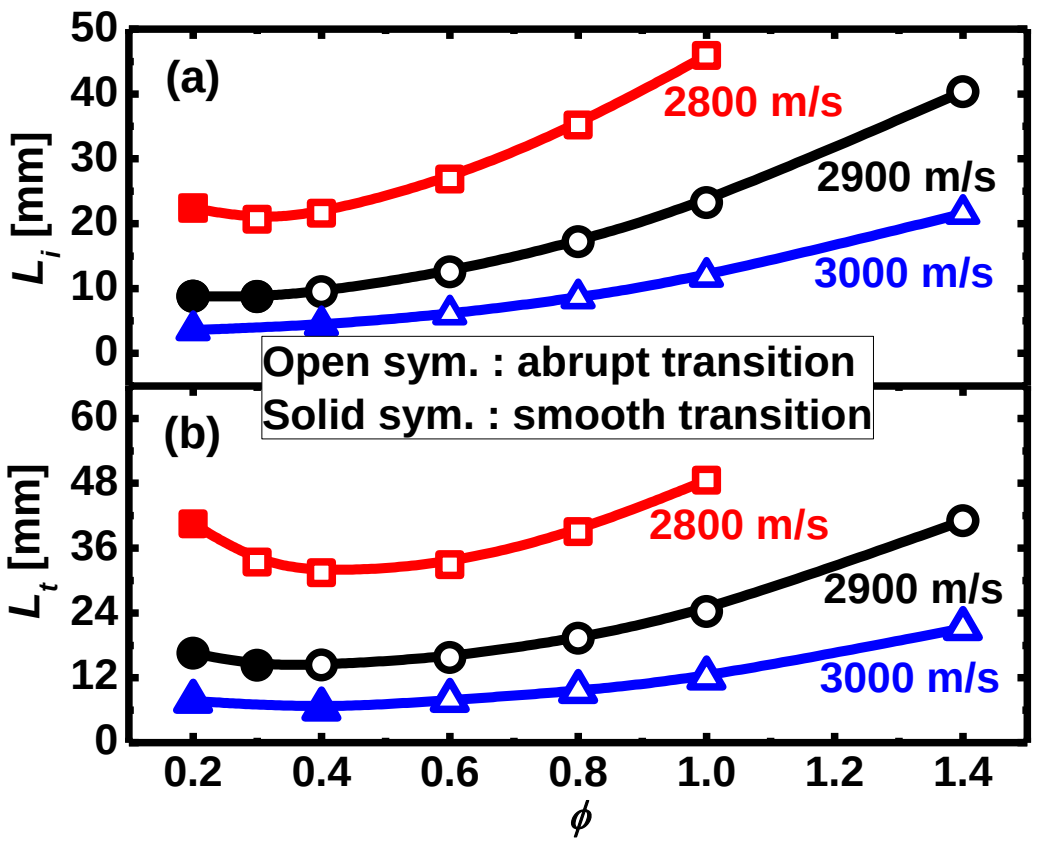

Fig. 5. Locations of (a) near-wedge initiation and (b) shock-to-detonation transition as functions of equivalence ratios with different freestream velocities.

Figure 5 summarizes the near-wedge initiation length $L_i$ and transition length $L_t$, predicted by ODW simulations, with different freestream equivalence ratios and velocities. As seen from Fig. 5(a), when $\phi$ is increased, $L_i$ monotonically increases (except for $\phi$ = 0.2, $u_{in}$ = 2,800 m/s). As shown in Fig. S7 of the supplementary document, for a fixed freestream velocity, the post-shock pressure exhibits slight increases, but the temperature almost linearly decreases with the equivalence ratio. For the curve of $u_{in}$ = 2,800 m/s, slight non-monotonicity is present for the ultra-lean mixtures, which may be due to slightly higher pressure for the case $\phi$ = 0.2. In addition, $L_i$ from the CFD is generally consistent with the product of post-shock gas velocity and ignition delay predicted by constant-volume homogeneous ignition (not shown in Fig. 5). This implies that the reaction initiation location of *n*-heptane ODW can be well predicted by 0D ignition calculations, as also observed for hydrogen and acetylene [3,14].

Notably, $L_t$ has a non-monotonic dependence on $\phi$, with the minimum $L_t$ at $\phi$ = 0.4 − 0.6 as demonstrated in Fig. 5(b). The left (right) branches of the $U$-shaped curves are approximately featured by smooth (abrupt) shock-to-detonation transition based on our simulations. Besides, with increased freestream velocity, e.g., from 2,800 to 3,000 m/s, both $L_i$ and $L_t$ decrease (closer to the wedge tip). This is because increased velocity leads to stronger precursor shock, and therefore higher temperature and pressure behind OSW.

## *3.3 Chemistry and shock interaction*

It is significant to understand the chemical reaction behind the precursor shock, particularly when the finite-rate chemistry is considered for a realistic fuel, like *n*-heptane. In this section, we use the chemical explosive mode analysis approach (see details in Ref. [21]) to determine the accurate reaction information in the induction zone of *n*-$C_7H_{16}$/air oblique detonations.

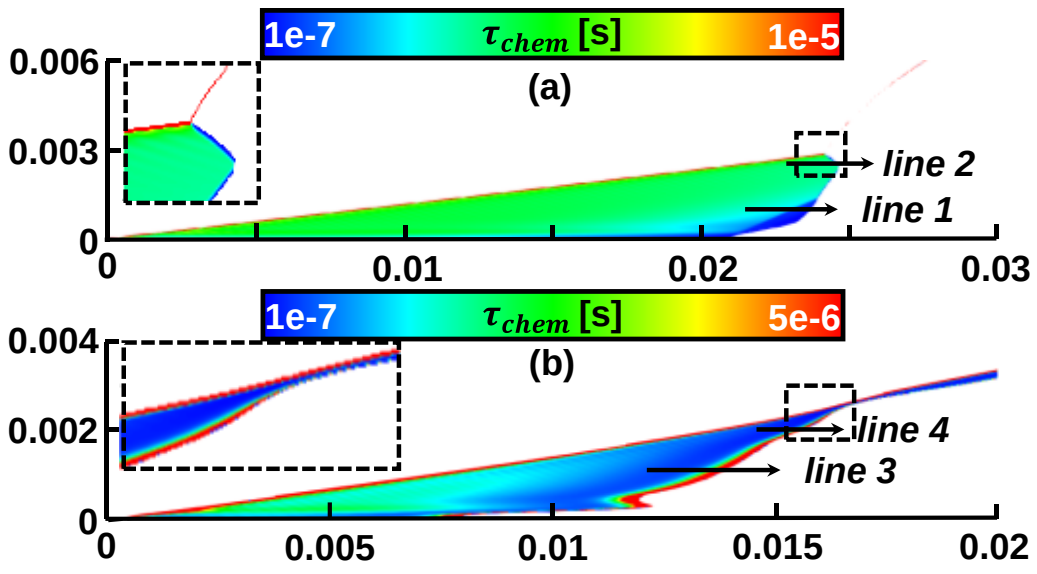

Fig. 6. Distributions of chemical timescale with equivalence ratio: (a) 1.0 and (b) 0.2. $u_{in}$ = 2,900 m/s. Axis label unit: m.

Figure 6 shows the distributions of chemical timescale $\tau_{chem}$ in two oblique detonations in Fig. 4. Here $\tau_{chem}$ is calculated from the reciprocal of the real part of the eigen value $\lambda_E$ for the chemical explosive mode [21]. Note that only chemical explosive areas with $Re(\lambda_E)$ > 0 is colored in Fig. 6. They essentially correspond to the induction zones between the precursor shock and reaction front (including deflagration and detonation waves). In both mixtures, after crossing the shock, the gas quickly changes from non-explosive to explosive state. Accordingly, $\tau_{chem}$ is gradually reduced, generally less than $10^{-6}$ s. In the induction zone, the local $\tau_{chem}$ in the stoichiometric *n*-heptane/air mixture ranges from $3\times10^{-6}$ s to $4.5\times10^{-6}$ s, which is longer than that in fuel-lean mixture (less than $2.5\times10^{-6}$ s). This is because that in the fuel-lean state, the gas temperature behind the oblique shock wave is relatively high.

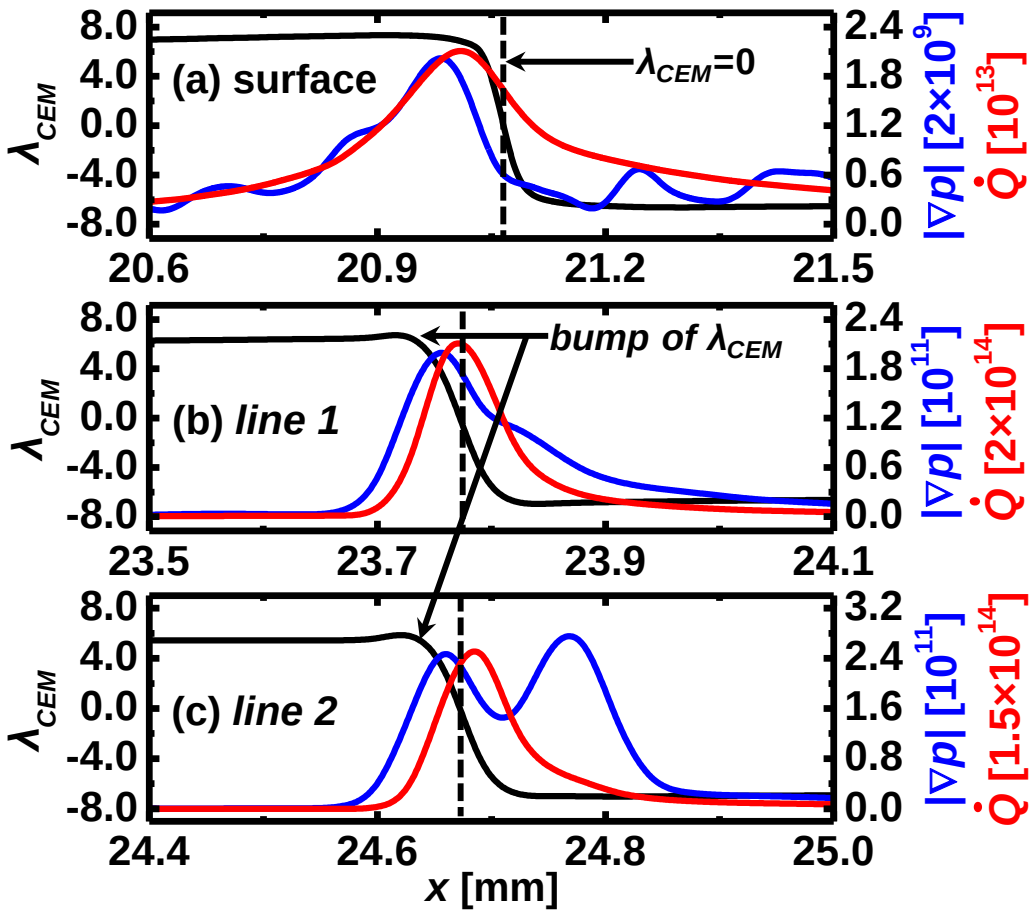

Fig. 7. Profiles of eigen value, pressure gradient magnitude (Pa/m) and heat release rate (J/m³s) with $\phi$ = 1.0. (a) line along the wedge surface $y = 1.0\times10^{-5}$ m, (b) line 1: $y = 1.0\times10^{-3}$ m, (c) line 2: $y = 2.5\times10^{-3}$ m.

The spatial evolutions of the eigen value of the chemical explosive mode, pressure gradient magnitude and HRR are shown for three locations: near the transition loci (lines 2 and 4 in Fig. 6), wedge surface, and middle of the reaction front (lines 1 and 3). The results of $\phi$ = 1.0 and 0.2 are given in Figs. 7 and 8, respectively. For clarity, the eigen value is visualized through $\lambda_{CEM} \equiv sign[Re(\lambda_E)] \cdot log_{10}[1 + |Re(\lambda_E)|]$. Note that $\lambda_{CEM} = 0$ (dashed lines) corresponds to a reaction front (RF), signifying the transition from unburned to burnt state [21].

According to Fig. 7(a), near the wedge, the HRR, $\dot{Q}$, increased quickly near the RF, accompanied by synchronously elevated pressure gradient magnitude $|\nabla p|$. This indicates that the reaction heat release generates strong shock from the ignition location due to the nature of isochoric combustion. Reaction-shock mutual enhancement also increases local gas reactivity, featured by the increased $\lambda_{CEM}$ around there. Further downstream in Figs. 7(b) and 7(c), the shock is significantly intensified ($|\nabla p|$ increases by almost 2 orders of magnitude), due to the continuous interactions between the shock and reactions. Near the multi-wave point in Fig. 7(c), a bump of $\lambda_{CEM}$ (marked with an arrow) can be seen. A detonation front is formed, which further intersects the precursor shock and hence the ODW is initiated.

Nonetheless, for the fuel-lean case (Fig. 8), the heat release is much slower in the induction zone, and the $\lambda_{CEM}$ reaches the maximum well head of the RF. Different from the results in Fig. 7, the interactions between the compressive waves and reactions are not observed. Based on our results, the chemical reactions inside the induction zone for this case is induced by the gradually curved OSW. The change of the precursor shock curvature starts at the location of $x$ = 50%$L_t$ or so and the OSW angle increases by about 18% [22], close to that (17.4%) from our case in Fig. 6(b). The precursor shock gradually compresses the post-shock mixture and therefore the reaction is initiated (see the variations of the chemical timescale in Fig. 6b). The distance between the OSW and the reaction zone is shortened gradually and finally the detonation wave is generated with a smooth mode.

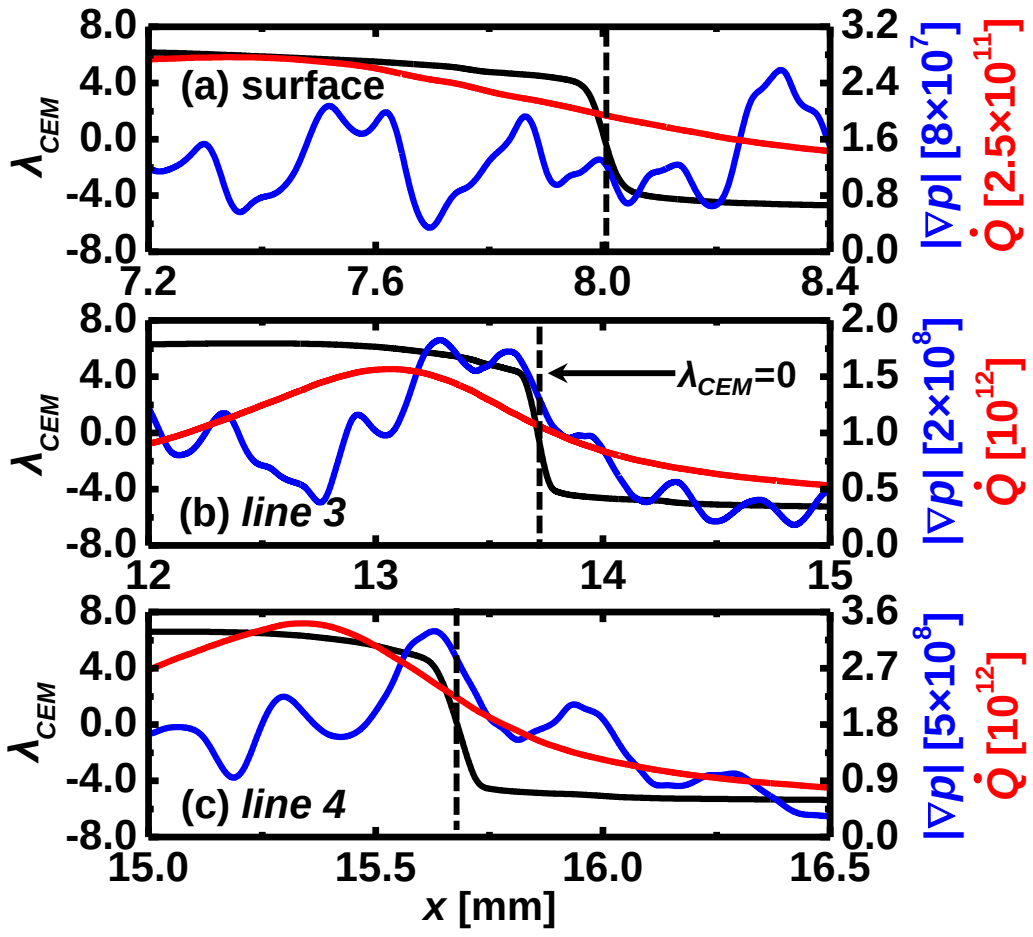

Fig. 8. Profiles of eigen value, pressure gradient magnitude (Pa/m) and heat release rate (J/m³s) with $\phi$ = 0.2. (a) line along the wedge surface $y = 1.0\times10^{-5}$ m, (b) line 3: $y = 1.0\times10^{-3}$ m, (c) line 4: $y = 2.0\times10^{-3}$ m.

### 3.4 *Predictability of transition mode*

Figure 9(a) quantifies the chemical excitation time $\tau_e$ and ignition delay $\tau_i$, to explore the possibility to predict the OSW-ODW transition mode with them. Both are from the 0D isochoric ignition calculations, taking the freestream conditions of our simulated cases in Fig. 5. The ignition delay $\tau_i$ is already defined in Section 3.1, while the chemical excitation time $\tau_e$ is the time duration from the instant of 5% maximum HRR to that of the maximum, which measures the rapidity of localized combustion heat release. This follows Bradley et al. [23]. Similar metrics to measure the chemical heat release time are also introduced by Figueira da Silva [11] and Morris [22], based on temperature rise. However, based on our study (see Fig. S8 in the supplementary document), temperature rise is not a general indicator to measure the heat release rapidity, particularly for fuel-lean mixtures.

One can see from Fig. 9 that, as the freestream velocity increases, both $\tau_e$ and $\tau_i$ decrease, but $\tau_i$ is more sensitive to the velocity variations. As seen in Fig. 9(a), under the same velocity, $\tau_i$ monotonically increases. This is because as the gas equivalence ratio increases, the temperature behind the OSW monotonically decreases. This leads to increased ignition delay $\tau_i$. However, since the pressure behind the OSW only has weak dependence of $\phi$ (see Fig. S7 of the supplementary document), reduced temperature (with $\phi$ ) leads to higher energy density of the combustible gas and therefore faster heat release [24]. This corresponds to smaller $\tau_e$.

Typically, rapid heat release measured by small excitation time $\tau_e$ would generate stronger pressure impulse from a local ignition spot, as indicated by Pan et al. [25]. Figure 9(b) summarizes the average pressure gradient magnitude at the initiation location (i.e., with peak HRR along the wedge surface) from our CFD simulations. They are compiled from 10 instants and the error bars denote the standard deviations. Together with Fig. 9(a), it is shown that the shorter excitation time of the gas composition correspond to higher pressure gradient magnitude around the near-wedge initiation location behind the OSW, indicating the emanating of stronger shocks from there. These strong shocks would further interplay with the chemistry in the induction zone as discussed in Section 3.3. Meanwhile, the pressure gradient magnitude of smooth ODW (fuel-lean cases, solid symbols in Fig. 9a) is generally lower than those of the abrupt ODW, indicating the good correlations between the excitation timescale, pressure wave intensity, and ODW initiation mode.

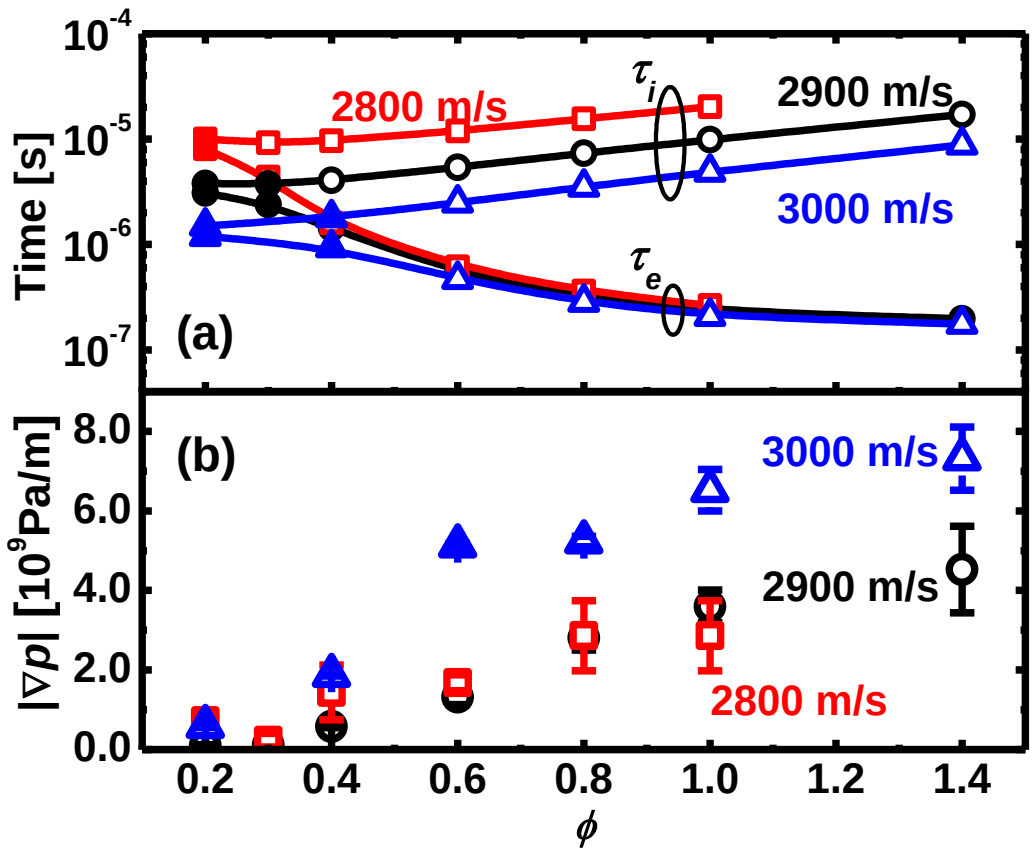

Fig. 9. (a) Ignition delay and chemical excitation time and (b) pressure gradient magnitude at the ignition location as functions of equivalence ratios and velocities. Open symbol: abrupt transition, solid symbol: smooth transition.

Figure 10 shows a diagram of chemical excitation time versus ignition delay. We add the scatters from our simulations in this work and from other numerical [3,4,14] and experimental [19,22] studies on oblique detonations with different fuels. Note that transition mode is differentiated with various symbols, which is determined from the simulation and experimental results. Here the two timescales are predicted from 0D calculations with the same chemical mechanisms as in the original references or with the high-pressure hydrogen mechanism [26] for the two experimental cases. We first look at our *n*-heptane results. For each freestream velocity, smooth transition occurs approximately between $\tau_e = \tau_i - 0.5\tau_i$ . This indicates that the time for heat release accounts for over 50% of the ignition delay, which implies that the heat release is slow. For acetylene ODWs [14], most of the smooth transition occurs around $\tau_e = 0.5\tau_i$. However, for hydrogen ODWs [3,4], smooth shock-to-detonation transition mostly occur when $\tau_e$ = 0.2-0.5$\tau_i$. Similar tendency is also observed from the ODW experiments [19,22]. Overall, we can see that when the ratio of chemical excitation time to ignition delay is relatively high (> 0.5 for *n*-$C_7H_{16}$, > 0.3 for $C_2H_2$ and > 0.2 for $H_2$, based on the existing compilation in Fig. 10), smooth transition is more likely to occur. The range of timescale ratio for more fuels and operating conditions should be determined from further studies. This may provide a approach towards theoretical or low-order engineering estimation for OSW-ODW transition mode.

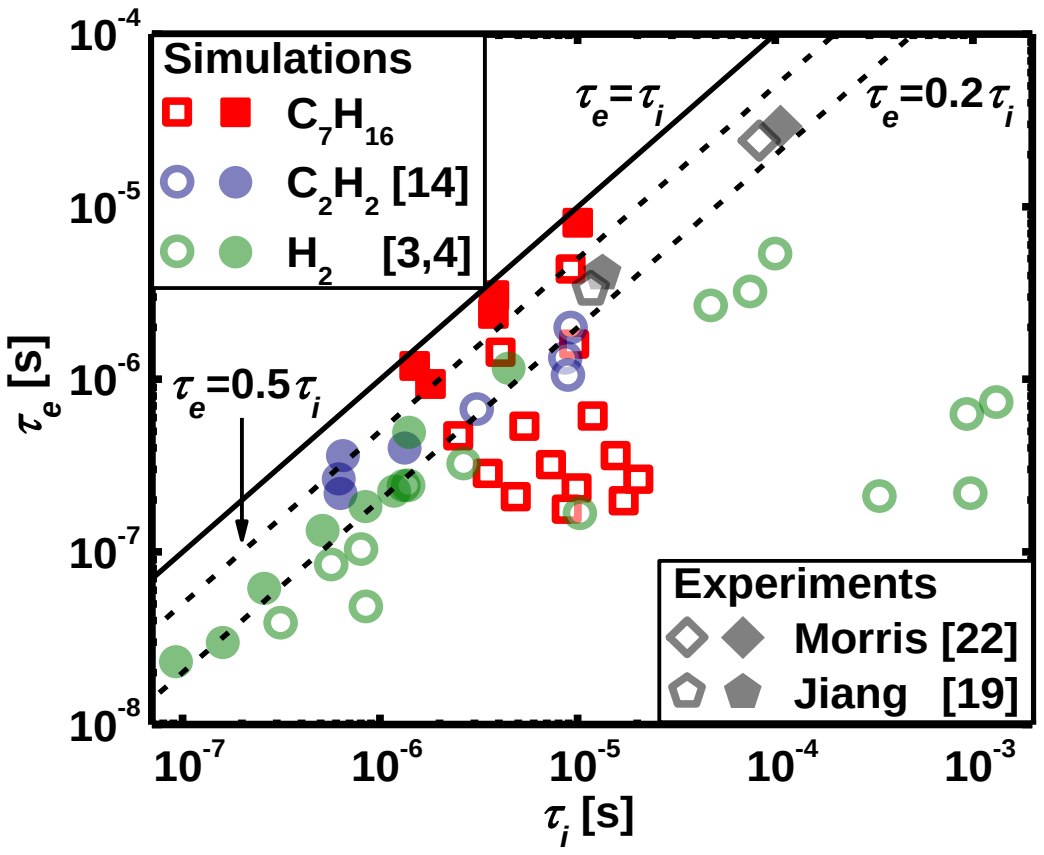

Fig. 10. Chemical excitation time versus ignition delay. Open symbol: abrupt transition, solid symbol: smooth transition.

## 4. Conclusions

Numerical simulations of oblique detonation wave in *n*-heptane/air mixtures are conducted, through solving the reactive Euler equations with a skeletal chemical mechanism. Different freestream equivalence ratios and velocities are considered, in which abrupt and smooth transition from oblique shock to detonation are predicted. Ignition limit, ODW characteristics, as well as mechanism and predictability of transition mode are discussed. The following conclusions can be drawn:

(1) Homogeneous ignition calculations are performed for both fuel-lean and stoichiometric mixtures. The results show that the ignition delay generally increases with the wedge angle. However, a negative wedge angle dependence is observed, due to the fuel NTC effects. The wedge angle range for successful ignition behind OSW decreases when the wedge length is reduced.

(2) For stationary ODWs, the induction length generally increases with the freestream equivalence ratio, but the initiation length exhibits non-monotonic variation. Smooth ODW always occurs for some fuel-lean conditions (equivalence ratio < 0.4). The interactions between shock / compression waves from and chemical reaction inside the induction zone are also studied.

(3) The predictability of shock-to-detonation transition mode is explored through the relation between ignition delay and chemical excitation time. It is demonstrated that the ignition delay (excitation time) increase (decrease) with equivalence ratio under the studied freestream conditions. Smaller excitation time corresponds to stronger pressure waves from the ignition location behind OSW. When the ratio of excitation time to ignition delay is high (> 0.5 for $n$-$C_7H_{16}$, > 0.3 for $C_2H_2$ and > 0.2 for $H_2$), smooth transition is more likely to occur.

## Acknowledgements

HG is supported by CSC Scholarship (No. 202006680013).